\documentclass[conference]{IEEEtran}
\IEEEoverridecommandlockouts
\usepackage{multirow}

\usepackage{algorithm}
\usepackage{algpseudocode}
\usepackage{amsmath}
\usepackage{balance}

\usepackage{listings}
\usepackage[dvipsnames]{xcolor}
\usepackage[breakable]{tcolorbox}

\usepackage{url}                        % before hyperref
\usepackage[hidelinks]{hyperref}        % load last
\usepackage[caption=false,font=footnotesize]{subfig}
\usepackage{placeins} % for \FloatBarrier
\usepackage{float}
\usepackage{booktabs} % use booktabs tables

\usepackage{xcolor}
\usepackage{marginnote}
\usepackage{color}
\definecolor{dkgreen}{rgb}{0,0.6,0}
\definecolor{gray}{rgb}{0.5,0.5,0.5}
\definecolor{mauve}{rgb}{0.58,0,0.82}
\definecolor{myblue}{RGB}{0, 102, 204}
\definecolor{myred}{RGB}{180, 0, 0}

\newcommand{\tab}[1]{Table~\ref{#1}}
\newcommand{\fig}[1]{Fig.~\ref{#1}}
\newcommand{\sect}[1]{\S\ref{#1}}

\newcommand{\alg}[1]{Alg.~\ref{#1}}
\newcommand{\eg}{\textit{e.g.}, } 
\newcommand{\ie}{\textit{i.e.}, }

\usepackage{amssymb} % for checkmark symbol

\begin{document}

\title{SQLMorph: Query Mutation and Fine-Grained Metrics for Text-to-SQL Evaluation}

\author{
	\IEEEauthorblockN{Mohammadhossein Malekpour, Mohamed Riahi, Maxime Lamothe, Amine Mhedhbi}
	\IEEEauthorblockA{Polytechnique Montréal \\
	\{mohammadhossein.malekpour, mohamed.riahi, maxime.lamothe, amine.mhedhbi\}@polymtl.ca}
}

\maketitle

\begin{abstract}
Text-to-SQL systems translate natural language queries into executable SQL, democratizing access to structured data. Despite recent advances driven by large language models (LLMs), evaluation remains a major bottleneck: 
public benchmarks fail to capture the complexity of enterprise schema, while building private evaluation sets is costly and nondeterministic, making evaluation results difficult to reproduce. 
To address this issue, we present SQLMorph, a framework for Text-to-SQL evaluation via query mutation. 
SQLMorph introduces two techniques to automatically generate and expand evaluation sets: Join Query Expansion (JQE), which systematically increases structural complexity through 
valid join additions, and Textual Query Augmentation (TQA), which generates controlled natural language perturbations to assess robustness to linguistic variation. JQE and TQA create targeted choke points to challenge specific system components. 
When applied to state-of-the-art systems, 
JQE increases query coverage and reveals accuracy degradation as the number of joins grows. 
Meanwhile, TQA shows that linguistic brittleness induced by heavy abbreviation can reduce accuracy by up to $17\%$.

Beyond evaluation sets, SQLMorph introduces a family of execution-level metrics that address the limitations of current binary measures, such as Execution Accuracy. 
We define Execution Precision (EXP) and Execution Recall (EXR) to quantify the fraction of correct and recovered results, respectively, and combine them via F1 for unified scoring. 
Our experiments show that these relaxed metrics enable fine-grained analysis of over- and under-prediction, revealing differences across systems that binary metrics obscure. 
Together, SQLMorph's query mutation and fine-grained metrics support debugging and better align Text-to-SQL evaluation practices with real-world deployments.
\end{abstract}


\section{Introduction}
\label{sec:introduction}

% 1. Broad progress on Text-to-SQL
Natural language interfaces to databases aspire to democratize data access. 
They allow users to express information needs in natural language (NL) and obtain answers directly from structured data. 
At the core of this vision lies the \emph{Text-to-SQL} task, which translates NL queries into executable SQL. 
While the task has witnessed remarkable progress~\cite{DBS-078}, the advent of large language models (LLMs) has substantially improved translation accuracy and spurred the development of new techniques. 
This momentum has, in turn, driven the rapid emergence of benchmarks such as Spider~\cite{yu-etal-2018-spider} and BIRD~\cite{li-2024-bird}, 
and enabled early, controlled production deployments.

% 2. Shortcomings of public benchmarks
Despite this progress, current evaluation practices remain far from capturing the complexity of real-world enterprise workloads. 
Enterprise queries often span multiple tables, exhibit intricate join structures, and refer to schema elements with domain-specific terminology and abbreviations. 
In contrast, public benchmarks typically feature short, structurally simple queries defined over intuitive schema names. 
This leads to an overly optimistic assessment of system capabilities and thereby distorts cross-system comparisons. 
Recent efforts such as the Beaver benchmark~\cite{chen-2024-beaver} seek to mitigate these limitations by introducing more complex schemas. 
Nevertheless, progress remains incremental and labor-intensive, advancing only through the development of new benchmarks one at a time. 

% 3. Challenges of private datasets
Evaluating Text-to-SQL systems on private datasets provides a more realistic alternative. 
Ultimately, the choice of a production Text-to-SQL system should depend on its performance on the target workload. 
However, constructing high-quality evaluation sets over private data is expensive, requiring substantial manual annotation, domain expertise, and increasingly, LLM-based automation. 
These processes are nondeterministic and difficult to control, making reliable and reproducible evaluation challenging. 
As a result, evaluation in both public and private settings remains fragmented, costly, and hard to scale. 

% 4. Limitations of existing metrics: Execution Accuracy (EX)
Evaluation methodology is further constrained by the metrics used to measure system performance. 
The dominant standard, \emph{Execution Accuracy} (EX), assigns a score of~$1$ when the predicted query's output relation exactly matches that of the ground-truth query and~$0$ otherwise. 
While intuitive, this binary measure collapses a wide spectrum of outcomes into a single judgment, failing to capture partial correctness or provide diagnostic feedback. 
A query that retrieves nearly all correct tuples but misses one condition is indistinguishable from one that returns an irrelevant result. 
Similarly, structural differences such as column reordering or the inclusion of irrelevant columns are penalized equally. 
Consequently, EX provides only a coarse, outcome-based view of correctness. 
This coarseness leaves evaluation misaligned with real-world requirements, where understanding the degree to which a system errs and identifying which components of a Text-to-SQL system succeed or fail are essential.

% 5. Core argument, gap.
These challenges stem from a common limitation: current evaluation practices remain static, 
relying on fixed benchmark queries in the form of NL--SQL pairs and on coarse binary metrics.  
To address these challenges, we reconceptualize Text-to-SQL evaluation as a dynamic process and introduce \emph{SQLMorph}, a new evaluation framework. 
\emph{SQLMorph} automatically generates 
valid query variants to 
create \emph{choke points} and stress test system components, 
and introduces new execution-level metrics that move beyond binary correctness. 
These metrics are fine-grained and intended for offline development and evaluation to support diagnosis of over- and under-prediction, 
where ground-truth queries and outputs are available.

\subsection{Contributions}
%Our main contributions are as follows:

\begin{itemize}
	\setlength{\itemsep}{0pt}
	\setlength{\parskip}{0pt}
	\setlength{\parsep}{0pt}

	\item \textbf{Join Query Expansion (JQE)} (Section~\sect{sec:join-query-expansion})\newline
	We introduce \emph{Join Query Expansion} (JQE), an algorithmic technique for increasing SQL complexity by expanding the set of joined tables in a query. 
	JQE performs semantic validation to ensure that each expansion introduces genuine additional reasoning complexity, for example, by requiring joins that cannot be inferred transitively. 
	It further applies diversity-aware pruning to maximize structural coverage across the generated variants. 
	The resulting queries remain executable while becoming substantially more challenging. 
	On BIRD, JQE doubles the average join degree and increases join cyclicity by an order of magnitude, leading to an Execution Accuracy (EX) drop of up to 20\% for state-of-the-art systems. 
	Overall, JQE stresses the SQL generation component along the dimension of join-intensive query construction.
	
	\medskip
	
	\item \textbf{Textual Query Augmentation (TQA)} (Section~\sect{sec:textual-query-augmentation})\newline
	We propose \emph{Textual Query Augmentation} (TQA), a method for modifying schema and natural-language naming to evaluate robustness to textual perturbations. 
	TQA systematically transforms schema elements or NL queries into less natural, abbreviation-heavy forms 
	(\eg \textit{WaterTemperature}~$\rightarrow$~\textit{WtTp}) while preserving semantics and executability. 
	The resulting variants remain semantically faithful yet expose substantial sensitivity to naming naturalness, with an Execution Accuracy (EX) drop of up to 17\% across systems. 
	TQA therefore probes linguistic robustness and context understanding, both of which are central to practical Text-to-SQL deployment.
	
	\medskip

	\item \textbf{Fine-Grained Execution Metrics} (Section~\sect{sec:relaxed-evaluation-metrics})\newline
	We introduce two fine-grained execution-level metrics, {Execution Precision} (EXP) and {Execution Recall} (EXR), to complement the standard {Execution Accuracy} (EX). 
	EXP measures the fraction of predicted tuples that are correct, while EXR measures the fraction of ground-truth tuples that are recovered. 
	Their harmonic mean, F1, provides a unified summary, while the individual metrics expose over-prediction and under-prediction separately. 
	Together, these metrics provide a more diagnostic view of system behaviour and reveal  differences between systems that EX alone obscures.
\end{itemize}

\noindent These contributions establish a framework for \emph{evaluation set construction through query mutation and fine-grained metrics}. 
To support reproducibility, we release the SQLMorph framework and all experimental scripts as open-source software.\footnote{\url{https://github.com/dais-polymtl/sql-morph}}

\section{Background}
\label{sec:background}

\begin{figure}[ht!]
	\centering
	\includegraphics[width=\linewidth]{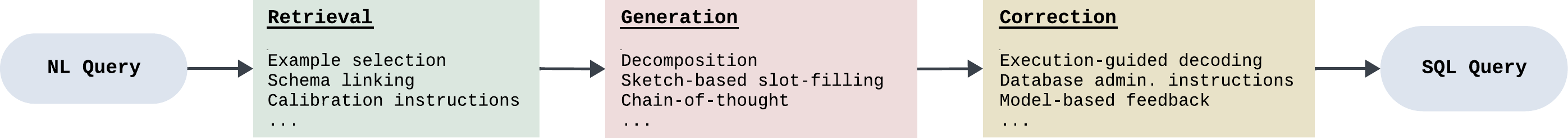}
	\caption{Text-to-SQL pipeline.} % TODO: hard to read.
	\label{fig:text-to-sql-pipeline}
\end{figure}

Modern Text-to-SQL systems typically rely on a multi-stage pipeline~\cite{hong2024nextgenerationdatabaseinterfacessurvey,Li2024TheDawn, liu2024surveynl2sqllargelanguage,zhang2024naturallanguageinterfacestabular,DBLP:journals/corr/abs-2406-12104}, as illustrated in \fig{fig:text-to-sql-pipeline}.  
A central design principle of \emph{SQLMorph} is to create targeted \emph{choke points} that systematically challenge specific components of this pipeline.  
Understanding these components is therefore essential to interpreting our evaluation results.

\subsubsection{Retrieval}
\label{subsubsec:retrieval}

This stage collects the contextual information needed to translate an NL query into SQL. 
This context may include schema elements such as tables, columns, and data types, representative database values, external documentation such as column descriptions, query-writing instructions, and demonstration examples~\cite{DINSQL,MACSQL,CHASE-SQL}.  
Retrieval methods are commonly divided into two categories: \emph{filtering}, which narrows the schema and candidate values to those most relevant to the query, and \emph{augmentation}, which enriches the prompt with additional useful context.  
Typical augmentation operations include entity extraction and the retrieval of explanatory passages through vector- or keyword-based search~\cite{CHESS,opensearchSQL}. 

\medskip

\subsubsection{Generation}
\label{subsubsec:generation}

Given the NL query and retrieved context, the generation stage has two main substeps: it first produces one or more candidate SQL queries and then selects a final output.  
Candidate generation methods include few-shot prompting, intermediate representations such as NatSQL~\cite{gan-etal-2021-natural-sql}, plan decomposition for complex queries, and sampling strategies that vary prompt templates or decoding temperature~\cite{n-rep,MCS-SQL,chainofthought}.  
The selection step then identifies the final SQL query using techniques such as self-consistency voting, rule-based filtering, and reranking.  
In practice, generation and selection are often interleaved in an iterative loop that runs for up to $k$ rounds, where feedback from the selection step guides the next round of candidate generation~\cite{XIYAN-SQL}. 

\medskip

\subsubsection{Correction}
\label{subsubsec:correction}

The correction stage aims to repair remaining errors in the selected SQL query.  
Common techniques include \emph{execution-based feedback}, which executes queries to identify issues such as syntax errors or empty results and then revises them accordingly, and \emph{model-based feedback}, which relies on auxiliary critic models or unit-test-style checks to detect and fix errors.  
Many systems apply correction iteratively, progressively refining plausible candidates into executable and semantically correct SQL queries~\cite{CSC-SQL}.

\medskip

Some systems include additional stages to refine generation through user feedback~\cite{DBLP:conf/cidr/MaamariLM25} or to reduce cost via LLM routing~\cite{DBLP:journals/corr/abs-2411-04319}. 
SQLMorph is designed to selectively target such stages through query-based mutations.
  
\section{Experimental Setup}
\label{subsec:experimental-setup}

This section outlines the experimental setup used to evaluate SQLMorph in  Sections~\ref{sec:join-query-expansion}--\ref{sec:relaxed-evaluation-metrics}. 
All of our experiments aim to reflect a realistic, medium-token-budget setting, with an explicit emphasis on minimizing cost for both generation and potential integration into CI/CD pipelines. 
We view our parameter defaults as a cost-conscious choice rather than a universally optimal setting, since the appropriate trade-off depends on the user's evaluation goals and resources.

\subsubsection{Dataset}
\label{subsubsec:dataset}
Our experiments use the BIRD benchmark~\cite{li-2024-bird}, a large-scale cross-domain dataset that has become a de facto standard for Text-to-SQL evaluation.  
BIRD contains 12,751 NL--SQL pairs across 95 databases spanning 37 domains.  
Each NL query may include \emph{evidence}, \ie external text that explains schema attributes, values, or computations, offering richer context for queries within specific domains.  

The dataset is divided into three splits: training (9,428 queries over 69 databases), development (dev) (1,534 queries over 11 databases), and hidden test (1,789 queries over 15 databases).  
We conduct all of our analyses on the dev set. 

\subsubsection{Models}
\label{subsubsec:models} 
SQLMorph includes a model manager component 
that supports a range of backends, including OpenAI, Hugging Face, and Ollama, enabling users to pin model versions and, when desired, to run local open-source models for reproducibility and to control model drift.
However, for the experiments in this paper, unless otherwise specified, we use \texttt{GPT-4o} and \texttt{text-embedding-3-small} from OpenAI as 
our models of choice. 

\subsubsection{Metrics}
\label{subsubsec:metrics}
For now, we adopt EX as the primary evaluation metric.  
EX compares the output relation of a predicted query \(\hat{R}_i\) with that of the ground-truth query \(R_i\), assigning a score of~1 for an exact match and~0 otherwise.  
The EX score on \(N\) queries is computed as follows: 
\[
EX = \frac{1}{N} \sum_{i=1}^{N} \mathbf{1}(R_i = \hat{R}_i) \; \text{where} \; \mathbf{1}(R_i, \hat{R}_i) = 
\begin{cases}
	1 & \text{if } R_i = \hat{R}_i \\
	0 & \text{otherwise.}
\end{cases}
\]

While this metric is standard, 
later sections (\sect{sec:relaxed-evaluation-metrics}) introduce SQLMorph's fine-grained alternatives that are relaxed to capture partial correctness and aid in error diagnosis. 

\subsubsection{Text-to-SQL Systems}
\label{subsubsec:text-to-sql-for-eval} 
We consider three representative open-source Text-to-SQL systems that differ in architecture as well as their retrieval and correction strategies. 
\tab{tab:system-performance} summarizes their reported performance on the BIRD leaderboard at the time of their submission. 
They are described below:
\begin{enumerate}
	\setlength{\itemsep}{0pt}
	\setlength{\parskip}{0pt}
	\setlength{\parsep}{0pt}
	\item[1.] {DIN-SQL}~\cite{DINSQL} 
	performs {schema retrieval} to align NL mentions with tables, columns, and values, then classifies queries as {easy}, {non-nested complex}, or {nested complex} to adapt prompting strategies.  
	It employs the intermediate representation {NatSQL}~\cite{gan-etal-2021-natural-sql} and applies a final self-correction stage via an LLM invocation.

	\item[2.] {MAC-SQL}~\cite{MACSQL} 
	follows the three-stage pipeline from Section~\ref{sec:background}, while adding a \emph{decomposer} to break complex NL queries into subproblems, generate subqueries for each during the generation stage, and then merge them.
	
	\item[3.] {CHESS}~\cite{CHESS}. 
	It augments the pipeline with an information retriever (\emph{IR}) that extracts keywords, values, and contextual hints from the NL query and schema.  
	Its candidate generator (\emph{CG}) repairs syntax errors, while a \emph{unit tester (UT)} executes candidates and selects the one passing the most validation checks.  
	We adopt CHESS\textsubscript{IR+CG+UT}, the strongest reported configuration.
\end{enumerate}

\begin{table}[t!]
	\caption{Performance as Execution Accuracy (EX) of three open-source Text-to-SQL systems on the BIRD benchmark at their time of submission.}
	\label{tab:system-performance}
	\centering
	\begin{tabular}{llrr}
		\toprule
		\textbf{System} & & \textbf{Dev} & \textbf{Test} \\
		\midrule
		Human (upper bound) & & -- & 93.0 \\
		CHESS\textsubscript{IR+CG+UT} & \cite{CHESS} & 68.3 & 71.2 \\
		MAC-SQL & \cite{MACSQL} & 57.6 & 59.6 \\
		DIN-SQL & \cite{DINSQL} & 50.7 & 55.9 \\
		\bottomrule
	\end{tabular}
\end{table}

This setup allows us to evaluate the three primary contributions of SQLMorph: Join Query Expansion, Textual Query Augmentation, and Fine-Grained Execution Metrics.

\section{Join Query Expansion}
\label{sec:join-query-expansion}

\subsection{Overview}
\label{subsec:jqe-overview}

Enterprise workloads frequently contain multi-table queries with complex join structures; yet public Text-to-SQL benchmarks underrepresent such patterns. 
\emph{Join Query Expansion (JQE)} within SQLMorph addresses this gap by {deterministically} increasing a query's structural complexity through the addition of
valid joins. 

Given an NL--SQL pair, JQE produces one or more {expanded} variants by adding a new table and  
valid join conditions. By iterating $n$ times over an evaluation set, JQE bootstraps queries with up to $n$ additional joins. 
Thus, applying JQE to an evaluation set can generate harder queries from only a small seed subset. 

\paragraph*{Intuition (running example)}
Suppose the original query retrieves customer names and total spending:
\begin{lstlisting}[language=SQL]
SELECT c.name, SUM(p.amount)
  FROM Customer c
  JOIN Purchase p ON c.id = p.cid
 GROUP BY c.name
\end{lstlisting}
If the schema contains 
\texttt{Purchase(pid,}~\texttt{cid,}~\texttt{sku)}, \texttt{Product(sku,} \texttt{price,} \texttt{category)}, 
and \texttt{Customer (id,} $\dots$\texttt{)}, 
JQE may expand the join query graph (JQG) by adding \texttt{Product} via \texttt{p.sku = Product.sku}, 
optionally exposing new attributes (\eg grouping by \texttt{category}) 
while maintaining the original intent and adding more information to the output. % producing a controlled variant. 
This stresses generation under larger join graphs (more tables, new predicates, and possibly cyclic JQGs) without resorting to  free-form LLM generation. 

\subsection{Design Principles}
\label{subsec:jqe-goals} 
Two principles guide JQE's design: (i) \emph{determinism and limited hallucination}: the core expansion primitive is algorithmic and graph-driven rather than generative, and LLMs are used only at the end to regenerate an NL query consistent with the expanded SQL; and (ii) \emph{user control}: users can constrain generation via diversity-aware pruning and preferences over join topology and \emph{coverage}. 
We define coverage in our join expansion context as introducing {new} join patterns, \ie 
JQGs that are non-isomorphic to those in an existing evaluation set.
Finally, we ensure that each added join is not redundant, \ie contributes genuine structural complexity. 
Specifically, the added join must not be implied through transitivity by the existing joins, and must therefore introduce at least one new explicit join condition in the SQL query.

\subsection{Approach}
\label{subsec:jqe-approach}
Our approach to JQE is divided into three parts:
\begin{figure}[t!]
	\centering
	\includegraphics[width=\linewidth]{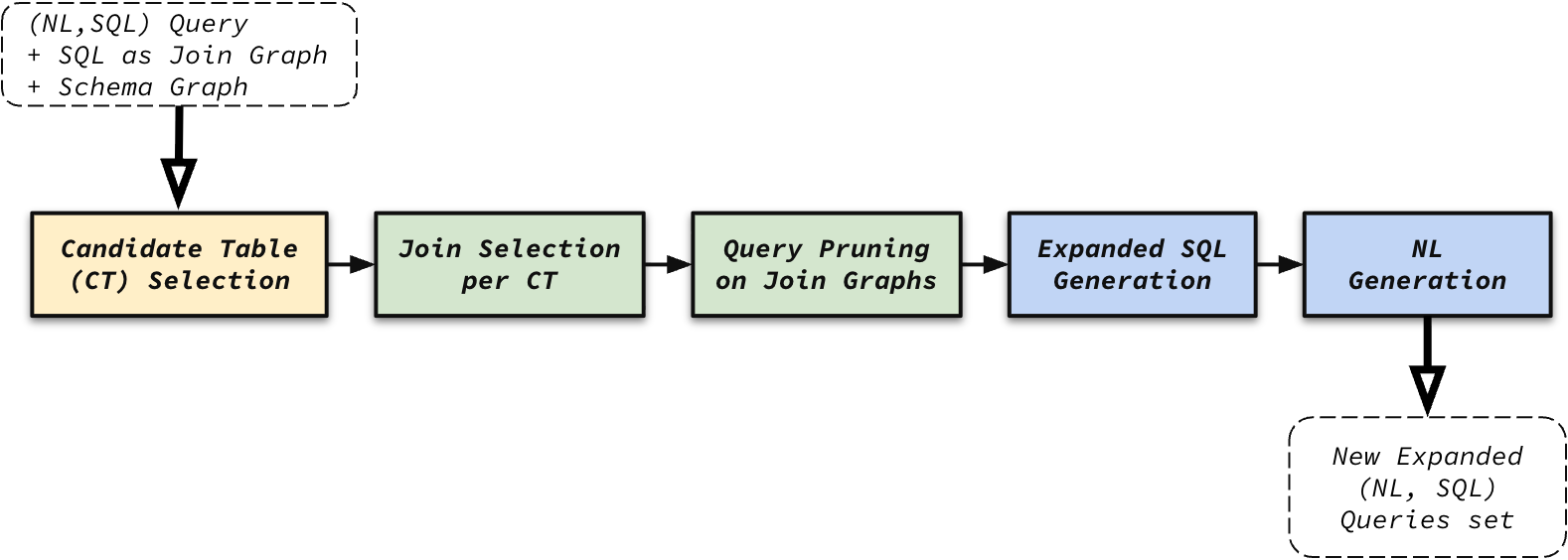}
	\caption{Multi-stage pipeline for join query expansion.}
	\label{fig:join_query_expansion}
\end{figure}

\subsubsection{Input}
\label{subsubsec:jqe-input}
\begin{itemize}
	\setlength{\itemsep}{0pt}
	\setlength{\parskip}{0pt}
	\setlength{\parsep}{0pt}
	\item \textbf{Schema graph.} 
	A graph $G_S=(V_S,E_S)$ whose vertices are tables and whose edges encode \emph{joinability}. 
	Each edge $e=(t_i,t_j)\in E_S$ carries one or more labels $L(e)$, where each $\ell\in L(e)$ denotes an admissible equi-join predicate (\eg $t_i.a = t_j.b$) derived from PK--FK constraints or other declared key compatibilities (\eg FK--FK joins).
	\smallskip
	\item \textbf{Seed query.} NL--SQL pair $(Q_{\text{NL}}, Q_{\text{SQL}})$ with $Q_{\text{SQL}}$'s join query graph $G_Q=(V_Q,E_Q)$, 
	where $V_Q\subseteq V_S$ are the tables appearing in \texttt{FROM} / \texttt{JOIN}, and $E_Q$ are the explicit join predicates.
	\smallskip
	\item \textbf{User preferences.}
	Optional parameters that bound and prioritize (ascending/descending) the expansion process:\vspace*{-0.25em}
	\begin{itemize}
		\item \emph{Join complexity preference:} whether to favour expansions that increase the number of join conditions per added table, or to restrict expansions to simpler, single-edge joins.  
		\item \emph{Centrality:} prioritization of expansions to more central tables in $G_S$ (\eg by degree or betweenness).  
		\item \emph{Query budget:} the maximum number of expanded NL--SQL pairs to generate.  
		\item \emph{Pattern coverage:} the maximum number of distinct, non-isomorphic join query graph (JQG) patterns to retain with reference to an input evaluation set.
	\end{itemize}
	By default, we prioritize higher complexity (more joins), arbitrary centrality, do not set a query budget, and set a maximum of a single unique JQG. 
\end{itemize}

\medskip

\subsubsection{Output}
A set of expanded NL--SQL pairs produced by augmenting the seed query with one table $t_e \in V_S \setminus V_Q$ and one or more admissible join predicates from $L(e)$. 
Each resulting variant is retained only if it satisfies the semantic and diversity constraints specified by the user preferences. 
The corresponding NL query is rewritten to reflect the added table and any newly projected attributes, with the aim of preserving the original query intent. 

\medskip

\subsubsection{Multi-stage pipeline}
\label{subsubsec:jqe-pipeline} 

\fig{fig:join_query_expansion} illustrates JQE as a multi-stage pipeline, and Algorithm~\ref{alg:jqe} specifies its operations. 
We briefly describe each stage below. 

\begin{itemize}
	\setlength{\itemsep}{0pt}
	\setlength{\parskip}{0pt}
	\setlength{\parsep}{0pt}
	\item \textbf{Stage A.} Candidate Table Selection.
	Using $G_S$, we enumerate candidate expansion tables, $v_e \in V_S \setminus V_Q$, that are adjacent to at least one $v_t \in V_Q$. 
	Each $(v_t,v_e)$ pair yields a potential join-condition expansion. 
	\item  \textbf{Stage B.} Join-Condition Combination (JCC) Generation.
	For each candidate table $v_e$, let $\{jc_k\}$ be the set of join labels on edges between $v_e$ and $\{v_t\in V_Q\}$. 
	We consider all non-empty subsets (\emph{power set}) of $\{jc_k\}$ as possible expansions for $v_e$, then prune combinations that introduce \emph{redundant} joins (defined below). 
	\item  \textbf{Stage C.} Diversity-Aware Pruning. 
	We rank JCCs using user preferences (\eg by added joins and centrality) and retain up to $n$ queries per \emph{unique} join pattern, where uniqueness is determined via JQG isomorphism (\texttt{IS\_UNIQUE\_JQG} in \alg{alg:jqe}). 
	This preserves structural variety while controlling cost.
	\item  \textbf{Stage D.} Expanded SQL Synthesis. 
	For each retained JCC, we programmatically inject the new table and join predicates into $Q_{\text{SQL}}$ (\eg by extending \texttt{FROM}/\texttt{JOIN} and \texttt{ON}/\texttt{WHERE} clauses).
	Optionally, simple predicates or projections may be added.
	\item  \textbf{Stage E.} NL Regeneration. A few-shot LLM prompt rewrites the NL query so that table/column mentions reflect the added join(s), while preserving the original intent. We keep prompts fixed for reproducibility. 
\end{itemize}

In summary, given $G_S$, JQE 
(A) identifies candidate tables for expansion, 
(B) enumerates all non-empty join-condition combinations (JCCs) for each candidate table,
(B$'$) discards JCCs with redundant joins, 
(C) performs diversity-aware pruning, 
(D) programmatically  synthesizes the expanded SQL query, 
and (E) regenerates the corresponding NL query via an LLM invocation. 
We next expand on redundancy checks and diversity-aware pruning. 

\begin{algorithm}[t]
	\caption{\textsc{expand\_joined\_tables}}
	\label{alg:jqe}
	\begin{algorithmic}[1]
		\State \textbf{Input:} schema graph $G_S$; NL--SQL pair $(Q_{\text{NL}}, Q_{\text{SQL}})$; $G_S=(V_S,E_S)$; preferences; \texttt{EvalSet}
		\State $\textsc{CTs} \leftarrow \emptyset$ // CTs: candidate tables for expansion
		\For{$v_t \in V_Q$} \Comment{Stage A}
		\For{$v_e \in \textsc{neighbors}_{G_S}(v_t)\setminus V_Q$}
		\State $\textsc{CTs} \leftarrow \textsc{CTs} \cup \{(v_t, v_e)\}$
		\EndFor
		\EndFor
		\State $\textsc{JCCs} \leftarrow \emptyset$ // join-condition combinations
		\For{$(v_t, v_e)\in \textsc{CTs}$} \Comment{Stage B}
		\State $\mathcal{J} \leftarrow \textsc{join\_labels}(v_t,v_e)$
		\For{$S \in \textsc{powerset}(\mathcal{J}) \setminus \{\emptyset\}$}
		\If{$\neg\,\textsc{has\_redundancy}(G_Q,S)$} \Comment{Stage C$'$}
		\State $\textsc{JCCs} \leftarrow \textsc{JCCs} \cup \{(v_e,S)\}$
		\EndIf
		\EndFor
		\EndFor
		\For{$(v_e,S) \in \textsc{rank}(\textsc{JCCs},\text{prefs})$} \Comment{Stage C}
		\State $Q'_{\text{SQL}} \leftarrow \textsc{inject\_joins}(Q_{\text{SQL}},v_e,S)$ \Comment{Stage D}
		\If{$\textsc{is\_unique\_jqg}(Q'_{\text{SQL}},\texttt{EvalSet})$}
		\State $Q'_{\text{NL}} \leftarrow \textsc{rewrite\_nl}(Q'_{\text{SQL}})$ \Comment{Stage E}
		\State $\texttt{EvalSet} \leftarrow \texttt{EvalSet} \cup \{(Q'_{\text{NL}},Q'_{\text{SQL}})\}$
		\EndIf
		\EndFor
	\end{algorithmic}
\end{algorithm}

\subsection{Semantic Redundancy Check}
\label{subsec:jqe-redundancy}
We remove redundant joins from the JQE-generated results. A join redundancy occurs when a candidate join is \emph{implied transitively} by existing joins (\eg $T_1.A=T_2.A$ and $T_2.A=T_3.A$ imply $T_1.A=T_3.A$). 
Such joins do not increase complexity and rarely appear in manually authored queries. 
We detect redundancy through \textsc{has\_redundancy} in \alg{alg:jqe} by examining cycles that include the candidate join expansion set, modeled as edges in the JQG; any JCC that forms such a transitive cycle is discarded. 

\subsection{Diversity-Aware Pruning via JQG Isomorphism}
\label{subsec:jqe-pruning}

We further refine the generated queries by pruning structurally equivalent variants. 
Two queries are considered \emph{structurally equivalent} when their JQGs are isomorphic~\cite{DBLP:journals/jacm/Ullmann76}. 
To allow flexibility, user preferences may specify that up to $n$ queries be retained for each unique JQG, with $n{=}1$ by default. 
We call the retained queries the \emph{expansion set} and the discarded queries the \emph{pruned set}; together, they constitute the \emph{generated set}. 

\subsection{Experimental Analysis}
\label{sec:expansion-evaluation}

To evaluate JQE, we follow the experimental setup in Section~\ref{subsec:experimental-setup} and apply JQE to the BIRD dev set with the default user preferences.

\medskip

\subsubsection{JQE Output Structural Distribution}
\label{subsubsec:structural-distribution-experimental-analysis}

JQE generates a total of $6{,}873$ queries (the \emph{generated set}). 
Among these, $58$ constitute the \emph{expansion set}, consisting of queries that are non-isomorphic to any query in the BIRD dev set or to any previously retained query, while the remaining $6{,}815$ are \emph{pruned} because they are isomorphic to a query in either the BIRD dev set or the expansion set already retained.
Overall, the generated set is $4.5{\times}$ larger than BIRD dev, and all queries execute with non-empty results. 
Table~\ref{tab:query_count} provides a summary.

\begin{table}[t!]
	\centering
	\caption{Query count in the sets: BIRD dev and JQE outputs.}
	\vspace*{-0.5em}
	\label{tab:query_count}
	\begin{tabular}{lcccc}
		\hline
		& \textbf{BIRD dev} & \textbf{Generated} &  \textbf{Pruned} & \textbf{Expansion} \\
		\hline
   		\textbf{Query Count} & 1{,}534 & 6{,}873 & 6{,}815 & 58 \\
   		\hline
	\end{tabular}
\end{table}

Since JQE rewrites NL queries after SQL expansion (Stage~E), we perform a lightweight human verification of NL--SQL pair alignment on the full expansion set ($58$ pairs). 
Each expanded NL--SQL pair is labelled as one of the following: \emph{fully captures the intent} (aligned; no edits needed), \emph{simply adds context/noise} (compatible but underspecified for the added join or columns), or \emph{loses the intent} (misaligned; would require correction). 
We find that 74.1\% ($43/58$) fully capture the intent, 13.8\% ($8/58$) simply add context or noise, and 12.1\% ($7/58$) lose the intent and require correction. 
This indicates that most rewritten NL queries remain faithful, while a small fraction exhibit underspecification or drift and require human or automated verification.

Next, we analyze the resulting structural distribution by comparing the average degrees and 
cyclicity of the JQGs in the BIRD dev set and the expansion set.

\noindent\emph{Join Query Graph Node Degree.} 
We measure connectivity using the average node degree of the JQG, $\bar d = 2|E|/|V|$. 
For BIRD dev, $\bar d{=}0.82$; for the generated set, $\bar d{=}1.35$; and for the expansion set, $\bar d{=}1.80$. 
Higher connectivity is driven by the default preference for greater join complexity, \ie more join conditions in each expansion. 

We further analyze per-query changes in the average node degree by comparing each original JQG ($G_{\text{orig}}$) with its expanded counterpart ($G_{\text{exp}}$). 
Across the $6{,}873$ generated queries, the most common degree increments are 
$\Delta \bar d{=}0.33$ ($3{,}910$ cases), 
$\Delta \bar d{=}0.17$ ($1{,}356$ cases), and 
$\Delta \bar d{=}1.00$ ($1{,}166$ cases). 
These distributions mirror the structure of the input queries: $58.7\%$ of the queries in BIRD dev originally join two tables, so adding a third table with a single join yields $\Delta \bar d{=}0.33$. 
Expansions that connect a new table to the two existing ones introduce higher connectivity, producing $\Delta \bar d{=}1.00$.

\noindent\emph{Join Query Graph Cyclicity.} 
Cyclic patterns are rare in BIRD dev (0.27\%) but increase to 4.31\% in the generated set and to 51.72\% in the expansion set of $58$ output queries (Table~\ref{tab:cyclic_distribution}). 
By maximizing the number of join conditions and thereby introducing cycles, JQE reflects the cyclic structures present in the schema graphs of several BIRD databases.

\begin{table}[t!]
    \centering
    \caption{The percentage of cyclic and acyclic join query graphs for the original BIRD queries and the generated and expansion sets.}  
    \label{tab:cyclic_distribution}
    \begin{tabular}{lrr}
        \toprule
        \textbf{Set} & \textbf{Acyclic (\%)} & \textbf{Cyclic (\%)} \\
        \midrule
        BIRD dev & 99.73 & 0.27 \\
        Generated Set & 95.69 & 4.31 \\
        Expansion Set & 48.28 & 51.72 \\
        \bottomrule
    \end{tabular}
\end{table}

\medskip

\begin{tcolorbox}[colback=blue!5!white,colframe=gray!75!black]
	\textbf{Takeaway.} JQE substantially increases structural diversity: in our experiment, it is able to generate $6{,}873$ queries with $58$ new join patterns, where the JQG average degree rises from $0.82$ to $1.35$, and cyclicity from $0.27\%$ to $4.31\%$. These effects are even more pronounced in the expansion set of $58$ final queries.
\end{tcolorbox}

\medskip

\subsubsection{Accuracy after JQE}
\label{subsec:jqe-ex}
We evaluate \textbf{CHESS}, \textbf{DIN-SQL}, and \textbf{MAC-SQL} (from Section~\ref{subsubsec:text-to-sql-for-eval}) on:
(i) BIRD dev set; 
(ii) $Q_{\text{ori}}$: the $30$ original queries that yield the $58$ expansions; and 
(iii) $Q_{\text{exp}}$: the $58$ expanded queries.
\tab{tab:EX-dev-BIRD} reports the EX of all three systems on each query set.

\begin{table}[t]
    \centering
\caption{EX of CHESS, DIN-SQL, and MAC-SQL on the full BIRD dev set, on the 30 original seed queries that produce JQE expansions ($Q_{ori}$), and on the 58 corresponding expanded queries ($Q_{exp}$).}    \vspace*{-0.5em}
    \label{tab:EX-dev-BIRD}
    \begin{tabular}{lccc}
        \toprule
        \textbf{System} & \textbf{EX\textsubscript{dev}} & \textbf{EX\textsubscript{$Q_{ori}$}} & \textbf{EX\textsubscript{$Q_{exp}$}} \\
        \midrule
        CHESS   & 64.86\% & 63.33\% & 44.83\% \\
        DIN-SQL & 59.11\% & 60.00\% & 32.76\% \\
        MAC-SQL & 55.90\% & 40.00\% & 39.66\% \\
        \bottomrule
    \end{tabular}
\end{table}

We define $\Delta\text{EX}=-1$ when a system changes from correct ($\text{EX}{=}1$) on an original query ($Q_{ori}$) to incorrect ($\text{EX}{=}0$) on its expanded version ($Q_{exp}$), $\Delta\text{EX}=1$ for the reverse, and $\Delta\text{EX}=0$ for no change. 
\tab{tab:delta_EX_exp} reports the distribution of $\Delta\text{EX}$: degradations ($\Delta\text{EX}{=}{-}1$) are more frequent than improvements  ($\Delta\text{EX}{=}1$) for CHESS and DIN-SQL, with DIN-SQL most affected (24.14\%), followed by CHESS (18.97\%).

\begin{table}[t!]
    \centering
    \caption{Distribution of changes in EX ($\Delta$EX) between original queries and their corresponding expanded queries for CHESS, DIN-SQL, and MAC-SQL.}
    \vspace*{-0.5em}
    \label{tab:delta_EX_exp}
    \begin{tabular}{lccc}
        \toprule
        \textbf{System} & \textbf{$\Delta$EX=1} & \textbf{$\Delta$EX=-1} & \textbf{$\Delta$EX=0} \\
        \midrule
        CHESS   & 8.62\%  & 18.97\% & 72.42\% \\
        DIN-SQL & 5.17\%  & 24.14\% & 70.69\% \\
        MAC-SQL & 17.24\% & 8.62\%  & 74.13\% \\
        \bottomrule
    \end{tabular}
\end{table}

\medskip

\subsubsection{Join Complexity vs.\ Performance}
\label{subsec:jqe-complexity}
\begin{figure}[t!]
	\centering
	\includegraphics[width=0.45\textwidth]{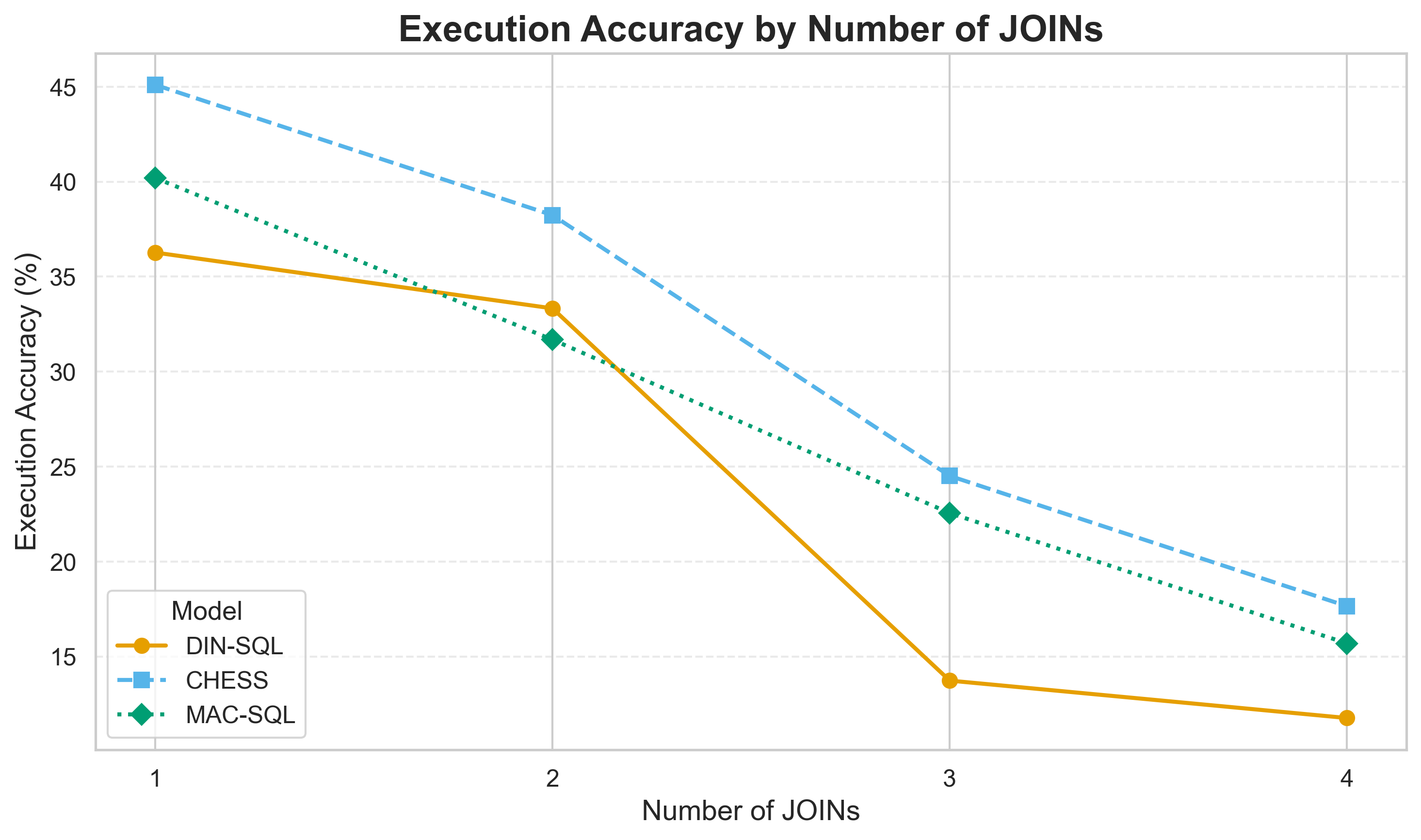}
	\caption{Execution accuracy (EX) decreases with the number of joins.}
	\label{fig:expanded-ex-per-joins}
\end{figure}

From the generated set, we sample four groups by JQG size: $(|V|,|E|)\in\{(2,1),(3,2),(4,3),(5,4)\}$. 
For each group, we keep the number of conditions and projections fixed, use the same number of queries per group ($120$, total $480$), and balance databases within each condition/projection combination. 
Figure~\ref{fig:expanded-ex-per-joins} shows EX changes across groups for the three systems. 
Accuracy declines monotonically with the number of joins, reaching $\leq\!15\%$ at higher join counts. 

Manual inspection attributes most failures to join-structure changes; a smaller fraction arise from secondary issues such as missing \texttt{DISTINCT}, projection errors, or \texttt{GROUP BY} mismatches. 
We ran an additional control experiment to better isolate the cause of the degradation. 
We use DIN-SQL, which is the most straightforward system to instrument 
because of its simple pipeline, and evaluate each seed query together with its corresponding expanded variant. 
Concretely, we first run DIN-SQL on the expanded NL query and capture the prompt of the generation stage. 
We then keep this context fixed, replace only the NL query with the original one, and regenerate the SQL. 
This experiment aims to disentangle the effect of join expansion (more involved query) from the effect of noise in the prompt by keeping the same context.

We report the transition rates among cases where the expanded NL query fails
($\text{EX}_{\text{exp}}{=}0$): Table~\ref{tab:din_control_conditional_zero} shows what fraction of these failures recover when we swap to the original NL under the same context, \ie $\Pr(\text{EX}_{\text{org}}{=}1 \mid \text{EX}_{\text{exp}}{=}0)$.
This recovery rate is substantial for small join sizes (\eg 54.90\% at $n_{\text{join}}{=}1$) and decreases as joins increase (down to 24.05\% at $n_{\text{join}}{=}4$), indicating that once the context is built for higher-join expansions, even the original NL becomes harder to solve under that fixed context. 
Despite this decreasing recovery rate, the consistent gains from expanded$\rightarrow$original across all groups still show that the additional join reasoning required by the expanded queries is a driver of the observed degradation. 
Even when degradation is possibly attributable to increased context noise, it is a consequence of queries that require more joins and thus expand to larger schema and evidence.

\begin{tcolorbox}[colback=blue!5!white,colframe=gray!75!black]
\textbf{Takeaway.} JQE efficiently stresses SQL structure: across CHESS, DIN-SQL, and MAC-SQL, EX drops as join complexity increases, confirming that join-heavy queries remain a key failure mode.
\end{tcolorbox}

\begin{table}[t!]
    \centering
    \caption{DIN-SQL recovery under fixed expanded-query context. For cases where the expanded query is incorrect ($\text{EX}_{\text{exp}}{=}0$), we replace the expanded NL query with the original NL query while keeping the generation-stage context unchanged, and report the EX transition rates by join count $n_{\text{join}}$.} 
       \vspace*{-0.5em}
    \label{tab:din_control_conditional_zero}
    \setlength{\tabcolsep}{6.5pt}
    \renewcommand{\arraystretch}{1.12}
    \begin{tabular}{c|rrr}
        \toprule
        $n_{\text{join}}$ & \textbf{$0\!\rightarrow\!1$ (recover)} & \textbf{$0\!\rightarrow\!0$ (still fail)} & \textbf{Total Qs ($\text{EX}_{\text{exp}}{=}0$)} \\
        \midrule
        1 & 54.90\% (28) & 45.10\% (23) & 51 \\
        2 & 41.54\% (27) & 58.46\% (38) & 65 \\
        3 & 34.67\% (26) & 65.33\% (49) & 75 \\
        4 & 24.05\% (19) & 75.95\% (60) & 79 \\
        \bottomrule
    \end{tabular}
\end{table}

\section{Textual Query Augmentation}
\label{sec:textual-query-augmentation}

\subsection{Overview}
\label{subsec:tqa-overview}

Schema naming conventions exert a strong influence on Text-to-SQL performance. 
Yet enterprise databases frequently employ domain-specific terminology and abbreviations, whereas public benchmarks predominantly feature more natural names. 
To alleviate this issue, \emph{Textual Query Augmentation (TQA)} targets the \emph{robustness} of Text-to-SQL systems by systematically altering the natural language (NL) query and schema identifiers in evaluation sets while preserving semantics and executability. 

Within \emph{SQLMorph}, TQA serves as a controlled mechanism for testing the retrieval stage, which is known to be sensitive to naming conventions and lexical overlap. Prior studies establish an inverse relationship between naming \emph{naturalness} and model accuracy, categorizing identifiers into three levels: regular (N1), low (N2), and least (N3)~\cite{SNAILS}. TQA challenges the retrieval stage by automatically transforming benchmark data into less-natural variants that maintain the same semantics.

\begin{figure}[t!]
	\centering
	\includegraphics[width=\columnwidth]{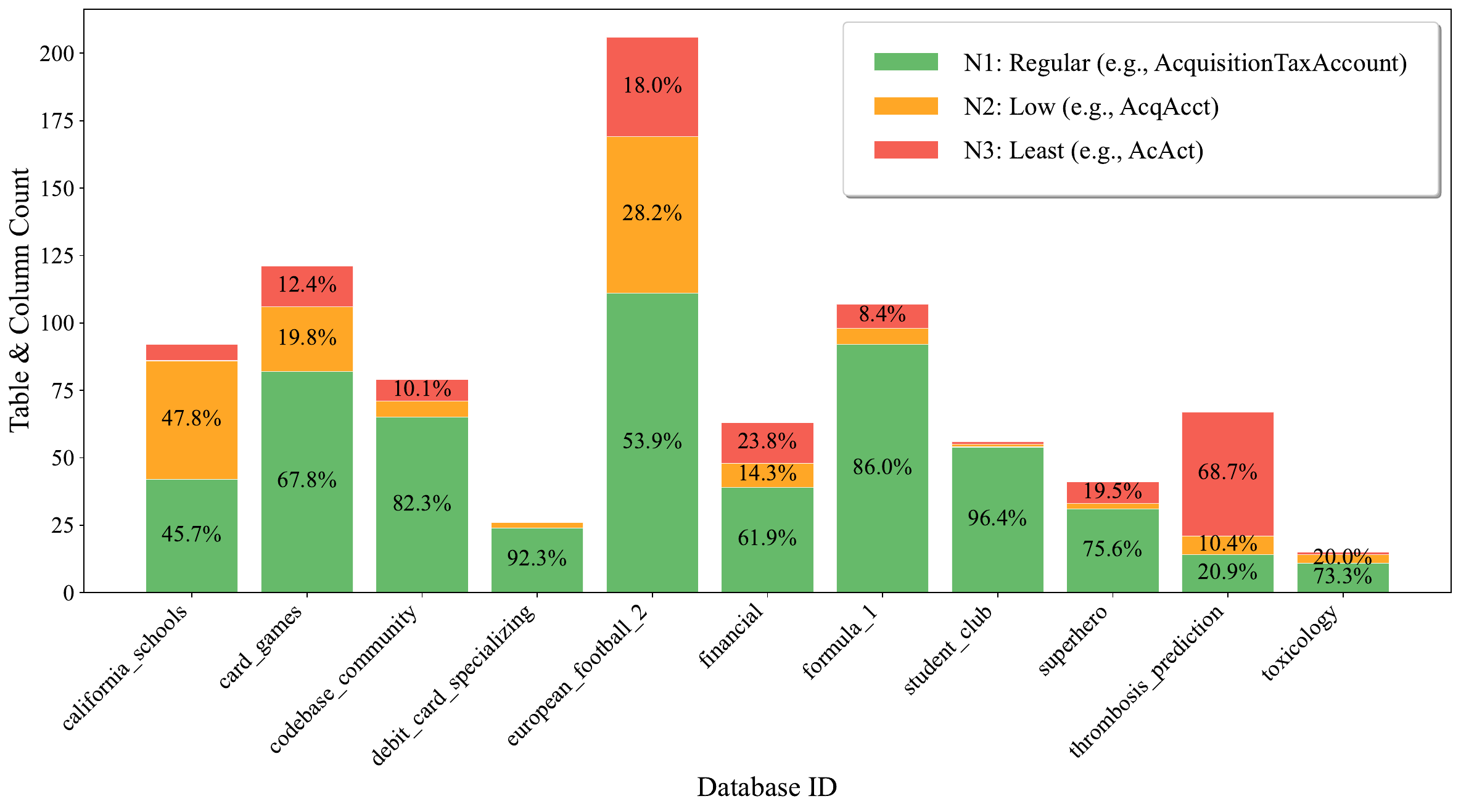}
	\caption{Distribution of naming naturalness across BIRD dev databases.}
	\label{fig:bird-naturalness-dist}
\end{figure}

Figure~\ref{fig:bird-naturalness-dist} illustrates the distribution of naming naturalness across the BIRD dev set, in which roughly 80\% of schema identifiers have natural (N1) forms. 
This imbalance motivates the need for augmentation techniques that can generate realistic, enterprise-like naming conditions to challenge retrieval stages and human-in-the-loop modules.

\subsection{Design Principles}
\label{subsec:tqa-principles}

One of TQA's design principles is to ensure that the observed performance changes arise solely from textual perturbations and not from semantic variation. 
As such, TQA transforms schema elements and their references within queries and evidence 
while keeping the SQL executable and replacing an element name in the SQL only when necessary. 
Values, constants, and clauses are kept fixed to guarantee minimal changes. 
This ensures that only lexical aspects of the query and not logic or intent, are modified.

\subsection{Approach}
\label{subsec:tqa-approach}
Our approach to TQA is divided into three parts:
\subsubsection{Input}
\label{subsubsec:tqa-input}

\begin{itemize}
	\setlength{\itemsep}{0pt}
	\setlength{\parskip}{0pt}
	\setlength{\parsep}{0pt}
	\item \textbf{Seed query.} NL--SQL pair with accompanying evidence and relevant schema.
	\smallskip
	\item \textbf{Naturalness classifier.} We use the SNAILS classifier~\cite{SNAILS}, 
	which was fine-tuned on \texttt{google/canine-s}, to assign each instance to one of three naturalness levels: regular (N1), low (N2), or least (N3). 
	\smallskip
	\item \textbf{User preferences (optional).} 
	De-naturalization aggressiveness (N1$\!\to$N3 vs.\ N1$\!\to$N2), and target references location: NL, schema, or both. 
	By default, we apply the most aggressive setting, targeting N3 in both the NL query and the schema.
\end{itemize}

\medskip

\subsubsection{Output}
\label{subsubsec:tqa-output}

A set of minimally changed, \emph{less-natural} references that remain semantically faithful and executable, potentially yielding 
(i) modified schemas (via \texttt{ALTER}); 
(ii) updated ground-truth SQL queries (with rewritten identifiers); 
and (iii) updated NL queries and evidence (with only schema mentions rewritten).
Each item passes execution equivalence checks (EX~$= 1$ relative to the original query).

\medskip

\subsubsection{Technique}
\label{subsubsec:tqa-pipeline}
TQA executes as follows: 

\begin{itemize}
	\setlength{\itemsep}{0pt}
	\setlength{\parskip}{0pt}
	\setlength{\parsep}{0pt}
	\item \textbf{Classify Naturalness.} 
	Extract table and column identifiers from the database schema and label each with the SNAILS classifier as N1, N2, or N3, together with a confidence score.
	\item \textbf{Decrease Naturalness.} 
	Transform N1 and N2 identifiers into N3, or into the user-specified target. Using an LLM invocation with temperature~0 and a fixed seed yields transformations such as \textit{WaterTemperature}$\rightarrow$\textit{WtTp}.
	\item \textbf{Alter Schema.} Clone each database and apply table and column renames via SQL \texttt{ALTER}. Validate the rename was successful 
	(\eg using \texttt{PRAGMA table\_info()}). 
	\item \textbf{Alter SQL Query.} 
	Update the ground-truth SQL query by replacing the old identifier using regex while avoiding SQL keywords, functions, and values. 
	We discard original NL queries for which the system did not obtain EX $=1$. 
	\item \textbf{Alter NL Query.} Rewrite the NL query and evidence by substituting only schema mentions with least-natural forms using a fixed few-shot prompt. We avoid changes to values (numbers, dates, entities). 
    By restricting edits to schema mentions, 
    this substitution preserves the original meaning and keeps the NL query semantically consistent. 
\end{itemize}

\subsection{Experimental Setup and Analysis}
\label{subsec:tqa-experiments} 
To evaluate TQA, we follow the experimental setup in Section~\ref{subsec:experimental-setup}. We evaluate four augmentation settings that rename either the NL side (NL query and evidence), the SQL side (schema definition and SQL query), or both: 
\begin{itemize}
	\item \textbf{O/O}: Original NL + Original SQL
	\item \textbf{L/O}: Less-natural NL + Original SQL
	\item \textbf{O/L}: Original NL + Less-natural SQL 
	\item \textbf{L/L}: Less-natural NL + Less-natural SQL
\end{itemize}

We apply TQA to the BIRD dev set (1,534 NL--SQL query pairs). 
We evaluate TQA at both the system and the stage levels, and we provide a qualitative example. 

\begin{table}[t!]
	\caption{Percentage change in EX ($\Delta$EX) on the BIRD dev set under the three TQA naturalness settings for CHESS, MAC-SQL, and DIN-SQL, relative to the original/original (O/O)}
	\vspace*{-0.5em}
	\label{tab:tqa-naturalness-res-ex}
	\centering
	\begin{tabular}{lccc}
		\toprule
		& \multicolumn{3}{c}{\textbf{$\Delta$EX (\%)}} \\
		\cmidrule(lr){2-4}
		\textbf{NL/Schema} & \textbf{CHESS} & \textbf{MAC-SQL} & \textbf{DIN-SQL} \\
		\midrule
		L/O & $-7.0$ & $-11.1$ & $-9.2$ \\
		O/L & $-6.7$ & $-9.7$ & $-23.4$ \\
		L/L & $-8.8$ & $-11.7$ & $-23.9$ \\
		\bottomrule
	\end{tabular}
		\vspace*{-1em}
\end{table}

\paragraph{System-level analysis} 
We consider only queries for which a system successfully generated a correct output, \ie queries for which O/O yields EX $=1$, and then measure changes under (L/O), (O/L), and (L/L). 
This leads to initial query counts per system as follows: 998 for CHESS, 857 for MAC-SQL, and 904 for DIN-SQL. 
Table~\ref{tab:tqa-naturalness-res-ex} reports $\Delta$EX (\%) relative to O/O (negative is a drop). 
All systems degrade consistently. 
Schema and NL de-naturalization both reduce EX; DIN-SQL shows the largest degradation, followed by MAC-SQL and CHESS. 
Thus, TQA can increase benchmark difficulty while preserving semantics. 

\begin{tcolorbox}[colback=blue!5!white,colframe=gray!75!black]
	\textbf{Takeaway.}
	TQA can make queries harder while preserving semantics and executability.
	We find that regardless of the augmentation setting, O/L, L/O, and L/L, all lead to degradation. 
	For production settings in which the schema cannot change, L/O provides a natural setting to challenge system retrieval. This further highlights the importance of query expansion and data expansion techniques in Text-to-SQL systems. 
\end{tcolorbox}

\paragraph{Stage-level analysis} 
To conduct a stage-level analysis, we adopt the schema retrieval formulation of prior work~\cite{maamari2024deathschemalinkingtexttosql} and consider three retrieval approaches:
\textbf{Full-Schema} (no filtering; retrieves over the full schema),
\textbf{TCSL} (Table-then-Column), and
\textbf{SCSL} (Single-Column). 
We evaluate them across O/O, O/L, L/O, and L/L on the augmented queries. 
We report two complementary retrieval metrics: \textbf{FPR} (\emph{False Positive Rate}; lower is better) and \textbf{SLR} (\emph{Schema Linking Recall}; higher is better). 
FPR quantifies the proportion of retrieved columns that are irrelevant to the query, thereby measuring retrieval noise. 
Lower FPR indicates a more precise retriever that introduces less distracting context for downstream generation. 
SLR, in contrast, measures whether all required columns are retrieved for each query. 
Results are summarized in Table~\ref{tab:schema-linking-results}.

\begin{table}[t!]
	\centering
	\caption{Schema retrieval performance across naturalness settings. FPR (\%): False Positive Rate (lower is better). 
	SLR (\%): Schema Linking Recall (higher is better).}
	\vspace*{-0.5em}
	\label{tab:schema-linking-results}
	\renewcommand{\arraystretch}{1.1}
	\setlength{\tabcolsep}{6pt}
	\resizebox{\columnwidth}{!}{%
		\begin{tabular}{l *{6}{c}}
			\toprule
			& \multicolumn{2}{c}{\textbf{Full-Schema}} 
			& \multicolumn{2}{c}{\textbf{SCSL}} 
			& \multicolumn{2}{c}{\textbf{TCSL}} \\
			\cmidrule(lr){2-3} \cmidrule(lr){4-5} \cmidrule(lr){6-7}
			\textbf{NL/Schema} & \textbf{FPR} & \textbf{SLR} 
			& \textbf{FPR} & \textbf{SLR} 
			& \textbf{FPR} & \textbf{SLR} \\
			\midrule
			O/O & 90.1 & 99.5 & 29.9 & 15.0 & 13.7 & 27.4 \\
			L/O & 88.0 & 30.4 & 29.2 & 11.7 & 14.3 & 21.4 \\
			O/L & 90.1 & 96.4 & 34.1 & 9.8 & 16.4 & 25.4 \\
			L/L & 88.2 & 28.2 & 33.8 & 14.1 & 16.1 & 20.1 \\
			\bottomrule
		\end{tabular}
	}
		\vspace*{-1em}
\end{table}

\begin{itemize} 
	\item \textbf{L/O.} 
	Relative to O/O, SLR drops sharply for Full-Schema (up to $-69.1\%$) and more moderately for SCSL and TCSL ($-3.3\%$ and $-6.0\%$). 
	FPR decreases moderately, with a slight increase for TCSL.
	These changes point to worse retrieval quality and help explain the EX reduction. 
	\item \textbf{O/L.} SLR again decreases: $-3.1\%$, $-5.2\%$, and $-2.0\%$ for Full-Schema, SCSL, and TCSL, respectively, while FPR rises for SCSL and TCSL ($+4.2\%$ and $+2.7\%$). 
	\item \textbf{L/L.} Moving from L/O to L/L, the additional changes are comparatively small: SLR changes by $-6.2\%$, $+2.4\%$, and $-1.3\%$, accompanied by only slight increases in FPR. 
\end{itemize}

\paragraph{Qualitative Example} 
To illustrate the augmentation process and its downstream effects, we show an example from the \texttt{toxicology} database (ID~245) in the BIRD dev set. 
The original and less-natural versions differ only in schema naming and in its propagation to the NL and SQL, while preserving semantics and execution results. 
We next present the original and less-natural NL query, the evidence, and the required changes to the relevant schema. 
We then analyze the behaviour of schema retrieval on this query. 

\medskip
\noindent\textbf{Original NL:} 
\emph{What is the average number of bonds the atoms with the element iodine have?}\\
\textbf{Less-natural NL:} 
\emph{What is the average number of bnds the atms with the Elmt iodine have?}

\medskip
\noindent\textbf{Original Evidence:}\\
\emph{atoms with the element iodine refers to element = `i'; average = DIVIDE(COUNT(bond\_id), COUNT(atom\_id)) where element = `i'}\\[2pt]
\textbf{Less-natural Evidence:}\\
\emph{atms with the Elmt iodine refers to Elmt = `i'; average = DIVIDE(COUNT(b\_id), COUNT(atmId)) where Elmt = `i'}

\medskip

\noindent\textbf{Original Relevant Schema:}\\
\texttt{atom.atom\_id}, \texttt{atom.element}, \texttt{connected.atom\_ id}, \texttt{connected.bond\_id}

\medskip

\noindent\textbf{Less-natural Relevant Schema:}\\
\texttt{atm.atmId}, \texttt{atm.Elmt}, \texttt{conn.atmId}, \texttt{conn.b\_id}

\medskip
\noindent\textbf{Schema Retrieval.} 
Under the \textbf{O/L} and \textbf{L/O} settings, systems frequently fail to retrieve the correct columns, \eg by missing \texttt{atom\_id} or \texttt{connected.atom\_id}, which leads to downstream generation and execution errors. 

\begin{tcolorbox}[colback=blue!5!white,colframe=gray!75!black]
\textbf{Takeaway.} 
TQA can stress-test retrieval and query augmentation techniques, as well as system-level query reformulation and understanding. 
TQA can help identify misalignment failures when matching an NL query to a schema, \eg in the schema linking stage.

\end{tcolorbox}

\section{Fine-Grained Evaluation Metrics}
\label{sec:relaxed-evaluation-metrics}

\subsection{Motivation}
\label{subsec:metrics-motivation}

The standard EX metric is binary: the predicted and expected output relations either match exactly or they do not. As a result, it compresses a wide range of outcomes into a single bit and discards potentially important signal. 
Queries that recover most of the correct rows but miss a single boundary condition receive the same score as queries that return entirely irrelevant results. 
Likewise, structural differences such as harmless extra columns are penalized as severely as true semantic errors. 
Consequently, EX obscures partial correctness, limits error diagnosis, and provides little insight into how a Text-to-SQL system fails, whether through under-prediction or over-prediction of rows, columns, or values.

\paragraph*{Our proposal}
We introduce a family of \emph{fine-grained execution-level} metrics to quantify how a predicted result deviates from the ground truth, rather than only whether it matches exactly. 
At the core are two complementary measures: \emph{Execution Precision} (EXP), which captures the fraction of predicted cells that are correct, and \emph{Execution Recall} (EXR), which captures the fraction of ground-truth cells that are recovered. 
Their F1 score provides a compact summary, while separate reporting of EXP and EXR reveals whether errors arise primarily from over-prediction, reflected in lower precision, or under-prediction, reflected in lower recall.

\subsection{Approach}
\label{subsec:metrics-approach}

Given a ground-truth query $q$ and a predicted query $\hat{q}$, we execute both to obtain the column-name sets $cols(q),cols(\hat{q})$ and the row multisets $rows(q),rows(\hat{q})$. 
If $rows(q)$ and $rows(\hat{q})$ are identical over their full schemas, then $\mathrm{EX}=1$, and we set $\mathrm{EXP}=\mathrm{EXR}=\mathrm{F1}=1$. 
Otherwise, we proceed with relaxed matching, which first optionally matches the columns and then the rows and cells. 
Finally, the evaluator chooses whether to penalize extra predicted columns.

\medskip

\paragraph{Column Matching}
We compare only those columns that the evaluator deems matchable; SQLMorph offers three matching regimes:
\begin{itemize}
	\item \textbf{Exact-Column (EC).} Match by exact column name: $c_{\cap} = cols(q)\cap cols(\hat{q})$.
	\item \textbf{Semantic-Column (SC).} Build a textual descriptor for each column (name plus top-$k$ values), embed the descriptors, restrict to type-compatible pairs and domain values when applicable, and compute a maximum-weight bipartite matching. Keep only pairs above a similarity threshold (\eg $0.7$). 
	For a fixed embedding model, semantic-column matching is fully deterministic and always yields the same matching; changing the model, however, may change the result. 
	Since matching is performed only over output columns (typically fewer than 20), the cubic assignment cost is negligible in practice and independent of the number of output rows. 
	\item \textbf{No-Column (NC).} Skip column matching entirely; matching then relies only on row and cell comparisons.
\end{itemize}

\medskip

\paragraph{Row/Cell Matching}
Let $c_{\cap}$ denote the matched column set. We then match $rows(q)$ and $rows(\hat{q})$ as follows:
\begin{itemize}
	\item \textbf{Exact-Cell (EC).} Under EC or SC, project $rows(q)$ and $rows(\hat{q})$ onto $c_{\cap}$ and treat the resulting rows as multisets. Under NC, compare rows directly through their implicitly matched value sets similar to the EX computation. 
	For each unique matched row pattern $r$, let $f_q(r)$ and $f_{\hat{q}}(r)$ denote its multiplicities in $rows(q)$ and $rows(\hat{q})$, respectively. Then
	\[
	\begin{aligned}
		|\mathrm{MatchedRows}| &= \sum_{r} \min(f_q(r), f_{\hat{q}}(r)),\\
		|\mathrm{MatchedCells}| &= |\mathrm{MatchedRows}| \cdot |c_{\cap}|.
	\end{aligned}
	\]
	\emph{Intuition:} Credit is assigned only when two rows match exactly over the aligned comparison units, namely matched columns under EC or SC, or implicitly matched values under NC.
	
	\item \textbf{Partial-Cell (PC).} Under EC or SC, first project $rows(q)$ and $rows(\hat{q})$ onto $c_{\cap}$; under NC, compare rows through their implicitly matched value sets. Matching then proceeds in two phases:
	\begin{itemize}
		\item \emph{Phase 1: exact matching.} First, check for exact matches, as in EC. These exact matches are removed, and the remaining unmatched rows are passed to the second phase.
		\item \emph{Phase 2: partial matching.} For each remaining $p \in rows(\hat{q})$ and ground-truth $g \in rows(q)$, compute the fraction of equal cells as a similarity score:
		\[
		\mathrm{sim}(p,g) = \frac{\#\{i \in c_{\cap} : p_i = g_i\}}{|c_{\cap}|}.
		\] 
		We then greedily select the pair with the highest similarity, add the matching cells $\{i \in c_{\cap} : p_i = g_i\}$ to $\mathrm{PartialMatchedCells}$, and remove both rows from further consideration. 
		This process continues until all remaining $rows(\hat{q})$ have been considered or until no unmatched $rows(q)$ remain.
		We sort the rows and columns of the outputs of $q$ and $\hat{q}$ before matching so that equivalent relations for the same named attributes produce deterministic matches.
	\end{itemize}
\end{itemize}

Finally, the total number of matched cells, $|\mathrm{MatchedCells}|$, is  
$|\mathrm{ExactMatchedCells}| + |\mathrm{PartialMatchedCells}|$.

\medskip

\paragraph{Accounting for extra predicted columns}

We define $|cells(q)| = |rows(q)|\cdot |cols(q)|$ as the total number of ground-truth cells and account for extra predicted columns by choosing one of the following options:
\begin{itemize}
	\item \textbf{Penalize Extras (PE) predicted columns.} $|cells(\hat{q})| = |rows(\hat{q})| \cdot |cols(\hat{q})|$.  
	All predicted columns are counted, so extra columns reduce precision.
	\item \textbf{Ignore Extras (IE) predicted columns.} $|cells(\hat{q})| = |rows(\hat{q})| \cdot |c_{\cap}|$.  
	Only matched columns are counted, under the assumption that the predicted output already contains the required data and that extra columns may therefore be ignored. This option is unavailable under NC, since no matched column set $c_{\cap}$ is constructed. 
\end{itemize}

\subsubsection*{Metrics}
The final metrics are defined as follows: 
$$\mathrm{EXP} = |\mathrm{MatchedCells}| / |PCells|$$
$$\mathrm{EXR} = |\mathrm{MatchedCells}| / |GCells|$$
$$\mathrm{F1}  = 2 \cdot \mathrm{EXP} \cdot \mathrm{EXR} / ({\mathrm{EXP} + \mathrm{EXR}})$$

\begin{tcolorbox}[colback=blue!5!white,colframe=gray!75!black]
	\textbf{Takeaway.} EXP is the fraction of predicted cells that are correct, EXR is the fraction of ground-truth cells that are recovered, and $F1$ unifies the two. Depending on the evaluation setting, extra predicted columns may either be penalized or ignored.
\end{tcolorbox}

\subsection{Experimental Analysis}
\label{subsec:metrics-experimental-analysis}

\subsubsection{Controlled Error Sensitivity}

\begin{table*}[t!]
  \caption{Single-error mutations and their expected effects on the number of rows (R), columns (C), values (V), and evaluation metrics (EXP, EXR). Symbols denote the expected direction of change: $\uparrow$ increase, $\downarrow$ decrease, $=$ unchanged, and $\times$ context-dependent.}
  \vspace*{-0.5em}
  \label{tab:mutation_ops}
  \centering
  \resizebox{\textwidth}{!}{%
  \begin{tabular}{llll}
    \toprule
    \textbf{Operator} & \textbf{Description} & \textbf{(R, C, V)} & \textbf{(EXP, EXR)} \\
    \midrule
    \texttt{projection\_drop} & Remove one column from the projection list (if $>1$ remain). & $(=,\downarrow,=)$ & $(=,\downarrow)$ \\
    \texttt{add\_star\_wildcard} & Append \texttt{*} or \texttt{alias.*}/\texttt{table.*} if no star exists. & $(=,\uparrow,=)$ & (PE~$\downarrow$; IE~=,=) \\
    \texttt{distinct\_toggle} & Remove the \texttt{DISTINCT} keyword. & $(\times,=,=)$ & $(\downarrow,\times)$ \\\hline
    \texttt{where\_predicate\_delete} & Remove one predicate from a compound condition. & $(\uparrow,=,=)$ & $(\downarrow,\uparrow)$ \\
    \texttt{where\_condition\_flip} & Flip comparison direction ($=$$\leftrightarrow$$\neq$, $>$$\leftrightarrow$$<$, $\ge$$\leftrightarrow$$\le$). & $(\times,=,=)$ & $(\times,\times)$ \\
    \texttt{where\_strengthen} & Make boundary stricter (\texttt{<}$\rightarrow$\texttt{<=}, \texttt{>}$\rightarrow$\texttt{>=}). & $(\downarrow,=,=)$ & $(=,\downarrow)$ \\
    \texttt{where\_weaken} & Make boundary looser (\texttt{<=}$\rightarrow$\texttt{<}, \texttt{>=}$\rightarrow$\texttt{>}). & $(\uparrow,=,=)$ & $(\downarrow,\uparrow)$ \\
    \texttt{where\_remove} & Remove the entire \texttt{WHERE} clause. & $(\uparrow,=,=)$ & $(\downarrow,\uparrow)$ \\\hline
    \texttt{having\_condition\_flip} & Flip aggregate comparison ($=$$\leftrightarrow$$\neq$, $>$$\leftrightarrow$$<$, $\ge$$\leftrightarrow$$\le$). & $(\times,=,=)$ & $(\times,\times)$ \\
    \texttt{having\_remove} & Remove the \texttt{HAVING} clause. & $(\uparrow,=,=)$ & $(\downarrow,\uparrow)$ \\\hline
    \texttt{join\_break} & Remove the \texttt{ON} condition (cartesian product). & $(\uparrow,=,=)$ & $(\downarrow,\times)$ \\
    \texttt{join\_type\_to\_left} & Convert non-\texttt{LEFT} joins to \texttt{LEFT JOIN}. & $(\uparrow,=,=)$ & $(\downarrow,\uparrow)$ \\\hline
    \texttt{limit\_increase} & Double the \texttt{LIMIT} and add a random \texttt{OFFSET}. & $(\uparrow,=,=)$ & $(\downarrow,\uparrow)$ \\
    \texttt{limit\_decrease} & Halve the \texttt{LIMIT} (min 1). & $(\downarrow,=,=)$ & $(=,\downarrow)$ \\\hline
    \texttt{aggregation\_swap} & Swap aggregates (\texttt{SUM}$\leftrightarrow$\texttt{AVG}, \texttt{MIN}$\leftrightarrow$\texttt{MAX}, \texttt{COUNT}$\rightarrow$\texttt{SUM}). & $(=,=,\times)$ & $(\downarrow,\downarrow)$ \\
    \bottomrule
  \end{tabular}
}
\vspace*{-1em}
\end{table*}

\noindent\emph{Research Question} -- Can our fine-grained metrics capture and distinguish the effects of isolated single-error mutants in predicted SQL queries?

To evaluate SQLMorph's metrics, we construct a benchmark of single-error SQL mutants by applying each mutation to a ground-truth query and evaluating how the resulting change is reflected in the metrics. 
Each mutant is created by applying exactly one atomic change to a ground-truth query from the BIRD development set; the resulting mutations are summarized in \tab{tab:mutation_ops}. In the mutant-generation code, we enforce depth-1 mutations by exhaustively trying each mutation and retaining only those that both modify the AST and produce syntactically valid SQL. 
When a query contains subqueries, mutations are applied only to the outer query so that each change remains easy to interpret.

\tab{tab:mutation_ops} groups the mutants by SQL component and summarizes their expected effects on rows, columns, values, and our metrics. 
Some mutants mainly change the number of returned rows. 
For example, increasing the \texttt{LIMIT} or making a \texttt{WHERE} condition less restrictive tends to add rows, which should lower EXP. 
In contrast, decreasing the \texttt{LIMIT} or making a \texttt{WHERE} condition more restrictive reduces the rows, which should lower EXR. 
Other mutants affect columns or values. For instance, dropping a projected column should reduce recall, while adding a wildcard mainly hurts EXP when extra predicted columns are penalized. Mutants that alter aggregate values can reduce both EXP and EXR. 

\begin{table}[t]
  \caption{Impact of single-error mutation on evaluation metrics.}
    \vspace*{-0.5em}
  \label{tab:single-injected-errors}
  \centering
  \renewcommand{\arraystretch}{1.1}
  \setlength{\tabcolsep}{6pt}
  \resizebox{0.5\textwidth}{!}{%
  \begin{tabular}{lccccccc}
    \toprule
    & \multicolumn{7}{c}{\textbf{Evaluation Metrics}} \\ 
    \cmidrule(lr){2-8}
    & 
    & \multicolumn{3}{c}{\textbf{EC-EC-IE}} 
    & \multicolumn{3}{c}{\textbf{SC-EC-IE}} \\
    \cmidrule(lr){3-5} \cmidrule(lr){6-8}
    \textbf{Injected Error} 
    & \textbf{$\Delta$EX}
    & $\Delta$EXP & $\Delta$EXR & $\Delta$F1
    & $\Delta$EXP & $\Delta$EXR & $\Delta$F1 \\
    \midrule
    \texttt{projection\_drop} & -100 & 0 & -43 & -28 & 0 & -43 & -28 \\
    \texttt{add\_star\_wildcard} & -100 & -93 & -10 & -87 & -82 & 0 & -71 \\
    \texttt{distinct\_toggle} & -100 & -57 & -57 & -57 & -57 & -57 & -57 \\
    \addlinespace[3pt]
    \texttt{where\_predicate\_delete} & -100 & -85 & -60 & -81 & -85 & -64 & -82 \\
    \texttt{where\_condition\_flip} & -100 & -96 & -61 & -94 & -96 & -61 & -94 \\
    \texttt{where\_remove} & -100 & -96 & -54 & -95 & -96 & -54 & -95 \\
    \texttt{where\_strengthen} & -100 & -65 & -60 & -63 & -65 & -60 & -63 \\
    \texttt{where\_weaken} & -100 & -65 & -75 & -73 & -65 & -75 & -73 \\
    \addlinespace[3pt]
    \texttt{having\_condition\_flip} & -100 & -64 & -100 & -100 & -64 & -100 & -100 \\
    \texttt{having\_remove} & -100 & -61 & 0 & -51 & -61 & 0 & -51 \\
    \addlinespace[3pt]
    \texttt{join\_break} & -100 & -98 & -42 & -98 & -98 & -46 & -98 \\
    \texttt{join\_type\_to\_left} & -100 & -53 & -50 & -54 & -53 & -50 & -54 \\
    \addlinespace[3pt]
    \texttt{limit\_increase} & -100 & -76 & 0 & -62 & -76 & 0 & -62 \\
    \texttt{limit\_decrease} & -100 & -4 & -61 & -45 & 0 & -60 & -43 \\
    \addlinespace[3pt]
    \texttt{aggregation\_swap} & -100 & -100 & -100 & -100 & -100 & -100 & -100 \\
    \bottomrule
  \end{tabular}
  }
\vspace*{-1em}
\end{table}

\tab{tab:single-injected-errors} summarizes the results. 
We report $\Delta$EX, $\Delta$EXP, $\Delta$EXR, and $\Delta$F1. 
Smaller values below 0 indicate a larger penalty or drop from 1. 
The results show the main weakness of {EX}. For every mutant, EX drops to $0\%$. As a result, it cannot distinguish a mild error from a severe one. 
In contrast, EXP, EXR, and F1 spread the errors across a much wider range and can reveal how a query is incorrect.

\emph{Row-count mutants} behave as expected. With \texttt{limit\_ increase}, the query returns too many rows, so EXP drops sharply while EXR stays at $1$. 
With \texttt{limit\_decrease}, 
the query returns too few rows, so EXR drops while EXP remains nearly unchanged. This shows that EXP and EXR clearly separate over-prediction from under-prediction of rows.

\emph{Schema-related mutants} also produce clear patterns. \texttt{projection\_drop} removes output columns, which lowers EXR while leaving EXP unchanged. \texttt{add\_star\_wildcard} adds extra predicted columns, which mainly lowers EXP. Under SC matching, this penalty is smaller, suggesting that SC is more robust to harmless schema noise. \texttt{distinct\_toggle} lowers both EXP and EXR, showing as expected the effect of duplicates on the multiset comparisons of the results.

\emph{Filter and grouping mutants} show a range of severity. Mutants such as \texttt{where\_predicate\_delete} reduce both EXP and EXR. Bigger changes such as removing the \texttt{WHERE} clause or flipping it causes an even larger drop. 
For grouping-related mutants, removing the \texttt{HAVING} clause behaves like over-selection: EXP decreases while EXR stays high. By contrast, flipping a \texttt{HAVING} condition can drive EXR to zero.

\emph{Join mutants} have their own effects. \texttt{join\_break}, which creates a cartesian product, causes one of the largest EXP drops and also lowers EXR. \texttt{join\_type\_to\_left} produces a smaller and more balanced degradation, which is consistent with a milder semantic change.

Across mutants, {SC} usually matches or improves on {EC}, especially when the error introduces extra but semantically related columns. {PC} helps mainly when predicted rows are close to the correct ones but not identical, such as after \texttt{join\_break}; otherwise, its behaviour is similar to EC. 

\begin{tcolorbox}[colback=blue!5!white,colframe=gray!75!black] 
	\textbf{Takeaway.} EXP and EXR provide much more useful diagnostic signal than binary EX. They distinguish over-prediction from under-prediction, separate row and column errors, and better capture cases where outputs are partially correct.
\end{tcolorbox}

\medskip

\begin{table*}[t!]
\caption{System-level comparison of CHESS, DIN-SQL, and MAC-SQL on the BIRD dev set, restricted to the shared-failed queries where all three systems have EX $= 0$. 
Fine-grained metrics (EXP, EXR, and F1) under multiple matching configurations reveal partial correctness and differences in failure behaviour.}
\label{tab:full-metrics-system-level-comparison}
\centering
\renewcommand{\arraystretch}{1.1}
\setlength{\tabcolsep}{6pt}
\resizebox{\textwidth}{!}{%
\begin{tabular}{l c *{15}{c}}
\toprule
 &  & \multicolumn{15}{c}{\textbf{Evaluation Metrics (\%) -- BIRD Shared Failures Subset}} \\ 
\cmidrule(lr){3-17}
 &  & \multicolumn{3}{c}{\textbf{EC-EC-IE}} & \multicolumn{3}{c}{\textbf{EC-PC-IE}} & \multicolumn{3}{c}{\textbf{SC-EC-IE}} & \multicolumn{3}{c}{\textbf{SC-PC-IE}} & \multicolumn{3}{c}{\textbf{NC-PC}} \\
\cmidrule(lr){3-5} \cmidrule(lr){6-8} \cmidrule(lr){9-11} \cmidrule(lr){12-14} \cmidrule(lr){15-17}
\textbf{System} & EX & EXP & EXR & F1 & EXP & EXR & F1 & EXP & EXR & F1 & EXP & EXR & F1 & EXP & EXR & F1 \\
\midrule
% \multicolumn{17}{c}{\textbf{}} \\
\addlinespace[3pt]
\textbf{CHESS} & 0.00 & 28.76 & 19.66 & 18.44 & 29.06 & 19.85 & 18.82 & 47.80 & 16.62 & 15.14 & 51.44 & 19.11 & 17.82 & N/A & N/A & N/A \\
\textbf{DIN-SQL} & 0.00 & 29.01 & 21.12 & 18.42 & 29.09 & 21.25 & 18.51 & 63.04 & 19.54 & 17.18 & 66.26 & 22.05 & 19.71 & N/A & N/A & N/A \\
\textbf{MAC-SQL} & 0.00 & 31.09 & 16.07 & 14.63 & 31.34 & 16.30 & 14.84 & 54.47 & 13.18 & 11.91 & 56.97 & 15.10 & 13.82 & N/A & N/A & N/A \\
\bottomrule
\end{tabular}%
}
\end{table*}

\subsubsection{System-Level Comparison}

\noindent\emph{Research Question} -- Do granular metrics reveal finer distinctions in how systems fail when compared to EX?

We conducted a comparative analysis of three Text-to-SQL systems, namely, CHESS, DIN-SQL, and MAC-SQL, on the BIRD dev set. 
Among the $1,534$ total examples, we identified $410$ queries for which all three systems produced $EX = 0$, representing $\sim$26.7\% of the benchmark. 
These shared failures pose a particular challenge: EX offers no information for ranking systems or for analyzing their behaviour on failed queries, despite the potential partial correctness of the predictions. 
\tab{tab:full-metrics-system-level-comparison} summarizes the results.  

\begin{figure}[t!]
\centering
\includegraphics[width=\linewidth]{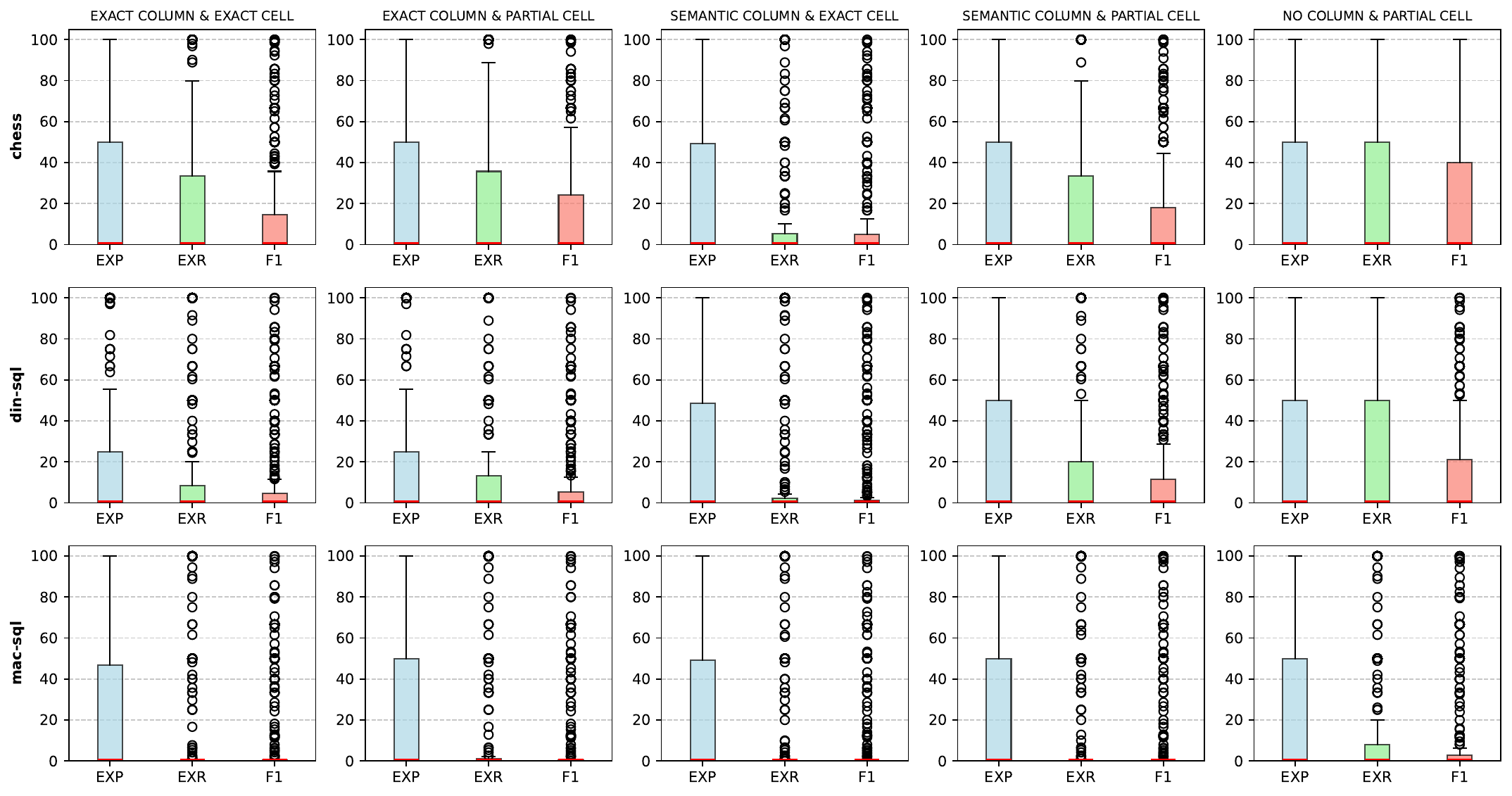}
\label{fig:failure-metrics-pe}
\vspace*{-1em}
\caption{System-level comparison on the BIRD dev set's common failed queries on combinations of fine-grained metrics.}
\label{fig:metrics-system-level-comparison}
\end{figure}

The ranking of the systems (CHESS $>$ DIN-SQL $>$ MAC-SQL) remains unchanged under both EX and our fine-grained metrics. 
However, EXP and EXR reveal meaningful differences within the shared-failure queries: for example, CHESS recovers nearly a quarter of the ground-truth cells, whereas MAC-SQL recovers only about 15\%. 
In addition, precision-recall asymmetries help explain the nature of these failures: CHESS maintains higher EXR, while DIN-SQL exhibits a more balanced trade-off between EXP and EXP. 
Thus, although the leaderboard order is stable, our metrics add diagnostic resolution.  
\fig{fig:metrics-system-level-comparison} illustrates this point by showing the distributions of EXP and EXR over failed queries, highlighting how close each system comes to the correct answer on average even when EX assigns them zero. 

\medskip

\subsubsection{Qualitative Example}

Consider the following NL query: `List the names of schools with more than 30 difference in enrollments between K--12 and ages 5--17. Please also give the full street address of the schools.'

\medskip

\noindent\textbf{SC-EC-PE Metrics (DIN-SQL vs MAC-SQL).}\\
\quad EX:\hspace*{1em} 0.000 vs 0.000 \\
\quad EXP:\hspace*{0.5em} 0.393 vs 0.500 \;(\emph{gap:} 0.11)\\
\quad EXR:\hspace*{0.48em} 0.982 vs 1.000 \;(\emph{gap:} 0.02)\\
\quad F1:\hspace*{1.4em} 0.561 vs 0.667 \;(\emph{gap:} 0.11)\\

This highlights how large the variation can be among failed queries. 
Next, we report the predicted and golden SQL queries and provide an explanation.

\vspace{2pt}
\noindent\textbf{Ground-truth SQL:}
\begin{lstlisting}[language=SQL]
SELECT T1.School, T1.Street
  FROM schools AS T1
  JOIN frpm AS T2 ON T1.CDSCode = T2.CDSCode
 WHERE T2.`Enrollment (K-12)` - T2.`Enrollment (Ages 5-17)` > 30;
\end{lstlisting}

\vspace{2pt}
\noindent\textbf{DIN-SQL Predicted SQL:}
\begin{lstlisting}[language=SQL]
SELECT frpm.`School Name`, schools.Street, schools.City, schools.Zip, schools.State
  FROM frpm
  JOIN schools
     ON frpm.CDSCode = schools.CDSCode
 WHERE (frpm.`Enrollment (K-12)` - frpm.`Enrollment (Ages 5-17)`) > 30;
\end{lstlisting}

\vspace{2pt}
\noindent\textbf{MAC-SQL Predicted SQL:}
\begin{lstlisting}[language=SQL]
SELECT T2.School, T2.Street, T2.City, T2.Zip
  FROM frpm AS T1
  JOIN schools AS T2 ON T1.CDSCode = T2.CDSCode
 WHERE T1.`Enrollment (K-12)` - T1.`Enrollment (Ages 5-17)` > 30;
\end{lstlisting}

\noindent\textbf{Explanation:}
\begin{itemize}
   \item Both systems predict the correct filtering condition for selecting schools with large enrollment differences, but they differ in their projected attributes.
  \item We would expect both systems to achieve an EXR of 1.0.
  However, DIN-SQL projects \texttt{frpm.`School Name'} instead of \texttt{School},  preventing a perfect match.
  \item DIN-SQL projects five attributes, whereas MAC-SQL projects four, corresponding to three and two extra attributes, respectively. This additional over-prediction lowers their EXP and, consequently, their F1 scores.
\end{itemize}

\medskip

\begin{tcolorbox}[colback=blue!5!white,colframe=gray!75!black]
\noindent\textbf{Takeaway:} This example shows high \textbf{EXR} but low \textbf{EXP}: both systems recover nearly all expected rows, but are penalized for projecting unnecessary columns. Binary EX treats them as equally wrong, whereas SQLMorph reveals that both are in fact very close to the correct answer. 
\end{tcolorbox}

\medskip

\section{Related Work and Challenges}
\label{sec:related-work}

We contextualize SQLMorph within recent research and emphasize persistent challenges in Text-to-SQL evaluation. 

Existing benchmarks such as \emph{Spider}, \emph{BIRD}, \emph{WikiSQL}, \emph{KaggleDBQA}, and \emph{Beaver} have driven substantial progress in Text-to-SQL research~\cite{yu-etal-2018-spider,li-2024-bird,qin2022surveytexttosqlparsingconcepts,zhong2017seq2sqlgeneratingstructuredqueries,lee-etal-2021-kaggledbqa,chen-2024-beaver}. However, they remain limited in scale, label fidelity, and compositional diversity. Public datasets also tend to emphasize relatively small schema and shallow join structures, which do not fully reflect the complexity of enterprise workloads~\cite{MitsopoulouK25}. Recent studies show that system accuracy drops sharply as join complexity increases~\cite{eckmann2025hlrsql}. SQLMorph complements these static benchmarks by generating deterministic query variants that increase both structural and lexical diversity. In particular, 
JQE introduces additional join predicates, thereby stressing systems along an underrepresented dimension in current benchmarks.

A related challenge is that real-world schema differ substantially from those in public datasets. Enterprise databases often contain hundreds of interrelated tables, heterogeneous sources, and cryptic identifiers~\cite{chen-2024-beaver}. 
Such complexity allows multiple valid SQL queries per NL intent~\cite{floratou2024nl2sql}, making correctness ambiguous even when intent is preserved~\cite{pourreza-rafiei-2023-evaluating}. 

Reproducibility remains another major concern. Progress can be difficult to interpret because results often depend on dataset splits, prompt design, sampling choices, and private evaluation settings~\cite{chen-2024-beaver,zhong2017seq2sqlgeneratingstructuredqueries,renggli2025evaluatingtext2sql,wenz2025benchpresshumanintheloopannotationrapid}. SQLMorph is designed to make evaluation more reproducible by fixing prompts, seeds, and sampling parameters, and by validating generated variants through execution-based checks. It is also suitable for private, in-house evaluation, where schema and workload traces cannot be shared due to privacy, governance, or annotation cost constraints~\cite{chen-2024-beaver,qin2022surveytexttosqlparsingconcepts}. In this sense, SQLMorph complements recent efforts on enterprise-oriented evaluation such as \emph{Beaver} and \emph{Spider 2.0}~\cite{chen-2024-beaver,lei-2024-spider2}. 

SQLMorph also relates to recent work on schema naming and schema linking. Prior studies have shown that naming quality strongly affects Text-to-SQL performance~\cite{SNAILS,qin2022surveytexttosqlparsingconcepts,katsogiannis2023survey,maamari2024deathschemalinkingtexttosql}. For example, \emph{SNAILS} emphasizes improving schema naturalness to facilitate linking~\cite{SNAILS}. 
SQLMorph takes a complementary perspective. TQA intentionally de-naturalizes schema identifiers and their mentions in natural language to simulate enterprise-style abbreviations and naming conventions. This allows us to stress-test robustness to lexical mismatch, schema linking, and retrieval stages more broadly. 

With respect to metrics, most existing leaderboards still rely on binary {Execution Accuracy} (EX)~\cite{yu-etal-2018-spider,li-2024-bird}. 
Although EX is simple and widely adopted, it collapses all failures into a single bit and does not indicate whether a system under-predicts rows, over-predicts rows, projects extra columns, or otherwise comes close to the correct answer~\cite{floratou2024nl2sql,kumar2022deep,pourreza-rafiei-2023-evaluating}. 
SQLMorph introduces {Execution Precision} (EXP) and {Execution Recall} (EXR), which separate over-prediction from under-prediction and provide more interpretable error diagnostics. 
SQLMorph adopts relaxed, execution-grounded metrics with cell-level matching, so that semantically equivalent or partially correct predictions can receive partial credit rather than being counted as complete failures.
Our metric design also differs from recent proposals based on continuous similarity scores~\cite{Pinna2025-jk}, which can be harder to interpret, and may conflate syntactic and semantic differences. 

Recent RL-based and reward-driven training methods for Text-to-SQL often rely on proxy, binary, or composite rewards that do not necessarily align with execution correctness at a fine-grained level~\cite{pourreza2025reasoningsqlreinforcementlearningsql,yao2025arctictext2sqlr1simplerewardsstrong,hao2025paverlsqltexttosqlpartialmatchrewards}. 
These rewards can reduce sparsity or improve alignment with end-task accuracy, but they do not explicitly decompose execution errors into over-prediction and under-prediction in the way that SQLMorph's EXP and EXR do~\cite{pourreza2025reasoningsqlreinforcementlearningsql,hao2025paverlsqltexttosqlpartialmatchrewards}. 
As such, 
SQLMorph's metrics may be worth exploring as potential training reward signals.

\section{Conclusion}
\label{sec:conclusion}

SQLMorph reframes Text-to-SQL evaluation as a form of controlled stress testing. 
JQE increases structural complexity through systematic join expansion, 
while TQA probes robustness to naming variation by de-naturalizing natural-language queries and schema references. 
Our fine-grained metrics, EXP and EXR, expose over-prediction and under-prediction that binary EX obscures. 
Together, these components provide a reproducible and diagnostic evaluation framework that better reflects enterprise requirements. 

\section{AI-Generated Content Acknowledgement}

We used GenAI tools to correct grammar and suggest synonyms or paraphrases for a few paragraphs. 

\balance

\bibliographystyle{IEEEtran}
\bibliography{references}

\end{document}